\documentclass[graybox]{svmult}

\usepackage{mathptmx}       
\usepackage{helvet}         
\usepackage{courier}        
\usepackage{type1cm}        
\usepackage{makeidx}         
\usepackage{graphicx}        
\usepackage{multicol}        
\usepackage[bottom]{footmisc}

\makeindex             
\newcommand{\apj}{ApJ}
\newcommand{\apjl}{ApJ}
\newcommand{\apjs}{ApJS}
\newcommand{\aap}{A\&A}
\newcommand{\aj}{AJ}
\newcommand{\mnras}{MNRAS}

\newcommand{\physrep}{Physics Reports}
\newcommand{\prd}{PRD}

\newcommand{\nat}{Nature}
\newcommand{\araa}{ARAA}
\newcommand{\jcap}{JCAP}

\newcommand{\na}{New Ast.}

\def\lsim{~\rlap{$<$}{\lower 1.0ex\hbox{$\sim$}}}
\def\gsim{~\rlap{$>$}{\lower 1.0ex\hbox{$\sim$}}}

\begin{document}

\title*{Cosmic Reionization and the First Nonlinear Structures in the Universe}
\author{Zolt\'an Haiman}
\institute{Zolt\'an Haiman \at Department of Astronomy, Columbia University, 550 West 120th Street, New York, NY 10027\\ \email{zoltan@astro.columbia.edu}}
%
%
\maketitle


\abstract{In this Introduction, we outline expectations for when and
  how the hydrogen and helium atoms in the universe turned from
  neutral to ionized, focusing on the earliest, least well understood
  stages, and emphasize the most important open questions.  We include
  a historical summary, and highlight the role of reionization as one
  of the few milestones in the evolution of the universe since the Big
  Bang, and its status as a unique probe of the beginning stages of
  structure formation.}

\section{Introduction}
\label{haimansec:1}

In the standard cosmological model, dominated by cold dark matter and
dark energy, the universe expands and cools dictated by the equations
of general relativity and thermodynamics, going through a handful of
global milestones.  Many of these milestones are well understood,
because their physics is within the reach of terrestrial experiments,
and observations leave little doubt about their occurrence.  These
begin with nucleosynthesis, and include the epoch of radiation-matter
equality, the recombination of hydrogen and helium, and the decoupling
of radiation.  In the more recent universe, dark energy has become
dominant and begun to accelerate the global expansion.

While the evolution of universe preceding nucleosynthesis is less well
understood, a generic prediction of inflation, describing the earliest
epochs, is the production of primordial density perturbations. These
perturbations obey Gaussian statistics, with a nearly scale-invariant
initial power spectrum.  The subsequent growth of perturbations over
time is again well understood, and leads to remarkable agreement with
many observations of the cosmic microwave background (CMB) and
large-scale structures (LSS).

The history of the universe is marked by additional mile-stones,
related to the growth of inhomogeneities. The first marks the epoch
when the first perturbations -- on astrophysically important scales --
reach non-linear amplitudes.  Ab-initio theoretical predictions become
more difficult at later epochs.  The first collapsed and
gravitationally bound structures form soon afterward, and serve as
the natural sites where the first stars and black holes then ``light
up'' the universe.  The reionization of the bulk of hydrogen and
helium atoms in the universe, several hundred Myr after the big bang,
is the most recent of these ``global'' milestones - resembling a phase
transition, and changing the character of the universe as a whole.

In addition to its fundamental place in our cosmic history, there are
three practical reasons why reionization is of special interest.
First, as will become clear below, and from later chapters in this
book, the bulk of reionization is thought to take place between
redshifts of $5\lsim z \lsim 10$. This range does not extend far
beyond our present observational horizon, and is within tantalizing
reach of experiments with forthcoming and planned instruments.  This
makes the study of reionization very timely.

Second, while there remains some room for more exotic scenarios,
reionization can be attributed to photo-ionizing radiation from two
different sources: an early generation of massive stars, or an early
generation of black holes powering (mini-)quasars.  The ultimate
energy source in these two scenarios is very different -- nuclear
binding energy, in the case of stars, and gravitational binding
energy, in the case of black holes.  These sources have different
efficiencies of producing radiation, and produce different spectra.
The details of how reionization unfolded thus depends on the
properties of these early stars and quasars (their luminosity and
spectral distribution) as well as on their abundance and spatial
distribution as a function of redshift.  

Finally, the earliest light-sources are quite plausibly too dim to be
detected directly, even with next-generation instruments. Studying
reionization is therefore one of the very few ways to glean details
about these first--generation objects.  It is worth emphasizing that
current observations only show the ``tip of the iceberg'': the
luminosity functions in even the deepest surveys show no evidence of a
faint-end turn-over, and we expect stars to form inside galaxies
orders of magnitude fainter than detectable at current and even
forthcoming flux limits.

In this article, we will first present a historical discussion of both
observations and modeling of the reionization history
(\S~\ref{haimansec:2}).  Then, in \S~\ref{haimansec:4}, we discuss two
possible ways to directly observe the light of the first generation of
ionizing sources.  It is important to emphasize that this article
contains a biased personal selection of some of the important
historical milestones and topics, and is not intended to be a
rigorous, complete review of the field.

\section{Historical Overview}
\label{haimansec:2}

\subsection{The Reionized IGM and its Observational Probes}

\subsubsection{Early History}

The realization that the mass density of neutral hydrogen (HI) in the
intergalactic medium (IGM) falls short by many orders of magnitude
quickly followed the identification of the first quasars in the early
1960s.  Here ``falls short'' is in comparison to the total mass
density expected from cosmology, i.e. comparable to the critical
density $\rho_{\rm crit}(z)=3H^2/8\pi G$, with $H=H(z)$ the
redshift-dependent Hubble parameter.  The quasar 3C~9 was among the
first handful of quasars discovered and identified through their
spectra.  At the time of its discovery, its redshift of $z=2.01$ was
an outlier, and held the distance record (with the other several
quasars at $z\lsim 1$)~\cite{Schmidt1965}.  Its spectrum lacked any
strong absorption on the blue side of the Lyman $\alpha$ emission
line, showing only a modest $\approx 40\%$ depression of the flux
instead ~\cite{GP65}.  This implies that the optical depth to Lyman
$\alpha$ scattering in the foreground IGM is $\tau_\alpha\approx 0.5$.
In their seminal paper, Gunn \& Peterson (hereafter GP) in
ref.~\cite{GP65} compared this to the optical depth,
$\tau_\alpha\approx{\rm few}\times10^6$ expected from Lyman $\alpha$
scattering by neutral hydrogen spread uniformly over the IGM, with a
near-critical mean density $\rho_{\rm crit}(z)\propto (1+z)^3$,
following the expansion of the universe.

It is worth quoting the result of this comparison: {\em ``We are thus
led to the conclusion that either the present cosmological ideas about
the density are grossly incorrect, and that space is very nearly
empty, or that the matter exists in some other form.''}  We now know
that the mean density of baryons is indeed lower than the critical
density, but ``only'' by a factor of $\approx$25.  We also know that
space can not be empty -- while large voids exist, their densities are
at most $\sim10\%$ below the mean.  The most plausible explanation, by
far, is that hydrogen is in ionized form. This was already the favored
interpretation at the time; however, interestingly, GP dismissed stars
and quasars as the primary ionizing sources. They instead considered
free-free emission or collisional ionization in the IGM itself, both
requiring that the IGM is hot ($\gsim 2\times 10^5$K).  Through the
study of the Lyman $\alpha$ forest, we now know that the IGM
temperature at $z \gsim 2$ is $T_{\rm IGM}\approx 10^4$K, more than an
order of magnitude lower than this lower limit.

Interestingly, GP already noted that a fully ionized IGM can produce a
large electron scattering optical depth.  Taking the electron (or
proton) number density from $\rho_{\rm crit}(z)/m_p$, this gives a
value of $\tau_e={\rm few}\times10\%$, which would be relevant for
observations of individual sources.  This is reduced by a factor of 25
by the low cosmic baryon density, to $\tau_e={\rm few}\times1\%$.

Remarkably, the cosmic microwave background (CMB) was discovered in
the same year in 1965; precisely 50 years
ago~\cite{PenziasWilson1965}.  This stimulated work on the
implications of the ionized IGM on the CMB.  With a hot IGM and a
large electron scattering optical depth, one would expect large
distortions in the spectral shape of the CMB
(e.g. ref.~\cite{SunyaevZeldovich1980,HoganKaiserRees1982}).  However,
the estimates of the baryon density and temperature were both soon
revised downward significantly.  Once again, observations of quasar
absorption spectra played important roles in these revisions.  First,
the discovery of the CMB also stimulated work on big bang
nucleosynthesis, making detailed predictions for the abundances of the
light elements.  The most important of these was the D/H ratio, which
placed a tight upper limit ($\Omega_b\equiv\rho_{b}/\rho_{\rm crit}\lsim
0.1$) on the baryon density.  Beginning in the mid 1990s, the relative
abundance [D/H] was measured in high-resolution quasar spectra and
resulted in the value $\Omega_b\sim 0.04$ (although less robust
non-cosmological D/H measurements pre-date these).  Second, as many
more quasars were discovered, and Lyman $\alpha$ absorption statistics
were collected over a large number of sight-lines, the modern view of
the Lyman $\alpha$ forest emerged.  This revealed that the low-density
IGM has a temperature of only $\sim 10^4$K, consistent with being
photoionized by the UV radiation of stars and quasars
\cite{Hernquist+1996,HaardtMadau1996}.

\subsubsection{Further Development of Observational Diagnostics}

In general, the highly ionized IGM can be studied either through
measurements of the residual neutral HI, or by detecting the effects
of the free electrons (and protons).  Beginning in the mid-1990s,
both of these possibilities were explored in great detail.

\vspace{\baselineskip}
\noindent{\em Effect of Free Electrons on the CMB.}

\noindent
On the ``electron side'', it was realized that even if the IGM is not
dense and hot enough to change the {\em spectrum} of the CMB, elastic
Thomson scattering by free electrons changes the patterns of both the
temperature and polarization power spectra (see reviews by
\cite{HaimanKnox1999,HuDodelson2002,Zaldarriaga+2008}).  Scattering a
fraction $\tau_e$ of the CMB photons out of each sightline translates
into a suppression of the primary CMB anisotropies (both temperature
and polarization) by a factor $\exp(-\tau_e)$, below angular scales
corresponding to the size of the cosmological horizon at reionization
(or $\lsim 10^\circ$ for reionization at $z\sim
10$)~\cite{HuWhite1997}.  This suppression can be difficult to
distinguish from a ``red'' tilt or a reduced normalization of the
primordial fluctuation spectrum.  However, scattering of the CMB
photons in the low-redshift ionized IGM also produces enhanced linear
polarization fluctuations on large scales (the so-called
``polarization bump'', ref.~\cite{HKR1982,Zaldarriaga+1997}). This
bump, on $\sim 10$ degree scales, is characteristic of reionization
and not present otherwise. The precise shape of this feature
(polarization power as a function of angular scale) can be used to
constrain the ionization history
\cite{Kaplinghat+2003,MortonsonHu2008}.  Finally, if reionization is
spatially inhomogeneous (patchy), as generally expected unless the
ionizing sources have unusually hard spectra, then this introduces
additional power on small ($\sim$ few arcmin) scales. Inhomogeneities
in the ionization fraction, rather than in the IGM density, can
dominate both the temperature and the polarization power spectra.
This was first shown in toy models \cite{GruzinovHu1998,Knoxetal1998}
and was later developed based on CDM structure formation models
(e.g. \cite{Santos+2003}; see ref. \cite{Mesinger+2012} for a recent
analysis of the kinetic Sunyaev-Zeldovich [kSZ] effect, which gives
the largest contribution).

As will be discussed in a later chapter in this book, the first
measurement of $\tau_e$ was made by the \texttt{WMAP} satellite, from
the temperature-polarization cross power spectrum, and yielded the
anomalously high value of $\tau_e\approx 0.17$ (translating to a
sudden reionization redshift near $z\sim 17$).  The increased
precision in subsequent \texttt{WMAP} measurements broke degeneracies
between $\tau_e$ and the spectral tilt $n_s$, and lowered this value
to $\tau_e\approx 0.08$.  The most recent determination from
\texttt{Planck}'s polarization power spectrum, $\tau_e\approx
0.066\pm0.016$~\cite{PlanckXIII}, remains consistent with this value,
and requires instantaneous reionization to occur around $z\sim
10$. More generally, the measured optical depth is twice the value
$\tau_e=0.04$ of the ``guaranteed'' contribution from the highly
ionized IGM between redshifts $0<z<6$. This requires that a tail of
ionization extends beyond the current observational horizon.  However,
such a tail is naturally expected even in the simplest models of
reionization, and leaves little room for additional, exotic ionizing
sources \cite{HaimanBryan2006,Visbal+2015b}.

\vspace{\baselineskip}
\noindent{\em Searching for Neutral Hydrogen.}

\noindent
Going back to history -- on the ``neutral hydrogen side'', work
continued on quasar absorption spectra.  An idea that dates back to at
least the early 1960s \cite{Field1962} is to detect intergalactic
neutral HI through its absorption in the 21cm hyperfine structure
line.  This ``radio analog'' of the GP trough, however, is much
weaker, due to the low oscillator strength of the 21cm line. As a
result, the corresponding upper limits on the neutral IGM density --
obtained from the lack of any 21cm absorption in the spectrum of the
$z=0.056$ radio galaxy Cygnus A \cite{Field1962} -- were $\sim 10^6$
times weaker than those obtained from the (lack of) Ly$\alpha$ GP
troughs.  Theoretical work on using the redshifted 21cm line, seen
either in absorption or emission (depending on the spin temperature)
in the context of an IGM being gradually ionized, and including
spatial fluctuations, dates back to ref.~\cite{HoganRees1979}.  The
idea apparently lay dormant for nearly two decades, but received
attention again from the mid-1990s, motivated by plans to build the
Giant Metrewave Radio Telescope (GMRT), and by the consensus emerging
about the modern CDM structure formation paradigm
\cite{SubramanianPadmanabhan1993,MadauMeiksinRees1997,Tozzi+2000}.  An
excellent review of the many ways of using the statistics of the
redshifted 21cm line to study reionization is given in
ref.~\cite{FurlanettoReview2006}.
 
In parallel with using the 21cm line, work continued on the utility of
the Ly$\alpha$ GP trough.  On the observational side, as more and more
distant quasars were discovered in the late 1990s, it became
increasingly puzzling that none of these showed the strong resonant GP
trough, expected even from a modestly neutral IGM. This was especially
so, since deep optical observations began to show that the abundance
of both quasars and galaxies decline beyond their peak at redshifts
$\sim 1-3$. The question arose whether the observed galaxies and
quasars can provide the required ionizing radiation -- it became
necessary to extrapolate well below the faint end of the observed
luminosity functions.

On the theoretical side, progress beyond the simple GP calculation of
the resonant optical depth $\tau_\alpha$, from a uniform IGM, was slow
to take off. However, beginning in the late 1990s, several studies
have begun to explore the expected absorption features in more detail.
For example, it was realized that the Ly$\alpha$ absorption from a
near-neutral IGM is so strong that the damping wings should be
detectable, and the red wings, in particular should offer a useful
diagnostic of a neutral IGM \cite{Jordi1998}.  Also, bright quasars
would be surrounded by a large (several Mpc) local ionized
bubble~\cite{ShapiroGiroux1987}, blue-shifting the observed location
of the GP trough and the damping wings \cite{CenHaiman2000}.  Another
realization was that there should be distinct absorption troughs at
Ly$\alpha$, Ly$\beta$, and possible higher Lyman lines, offering
another useful diagnostic~\cite{HaimanLoeb1999}, at least for the
first sources, that would be detected not far beyond the redshift
where the IGM turns predominantly neutral.  In the context of CDM
structure formation models, reionization must be gradual and
inhomogeneous, resulting in large line-of-sight
variations~\cite{MHR2000}.  All of the above effect had important
consequences once the first GP was discovered and had to be interpreted
(e.g. \cite{MH2004}).

The discovery \cite{Becker+2001} of the first GP trough was indeed a
watershed event in 2001.  The Keck spectrum of a $z=6.28$ quasar, one
of the first several $z\gsim 5$ quasars identified in the Sloan
Digital Sky Survey (SDSS), showed no detectable flux over a large
wavelength range short-ward of $\sim(1+z)1215$\AA.  This raised the
tantalizing possibility that 35 years after the seminal GP paper, we
have finally probed the era when the IGM was significantly neutral.
This discovery also stimulated a large body of work on the limits that
can be placed on reionization, given a ``deep'' and ``long'' dark
region (or regions) in the spectrum (e.g.~\cite{Barkana2002}).  The
issue is that ``zero flux'' can be consistent with resonant absorption
from the residual HI in a highly ionized foreground IGM.  Placing
constraints on reionization therefore necessitated detailed modeling
of the fluctuating IGM with a few Mpc of the quasar, including the
quasar's own ionized bubble.

Quasars are of course not unique -- a significantly neutral IGM would
imprint GP absorption features on any background source at
$\lambda_{\rm obs}=(1+z)\lambda_\alpha$.  It had long been expected
that a strong Ly$\alpha$ {\em emission} line would be produced by the
first ``primeval'' galaxies \cite{PartrigePeebles1967}. Numerous
searches for high-redshift galaxies using their Ly$\alpha$ emission,
however, did not yield any discovery for $\approx$ two decades -- the
failure was blamed on extinction of this line by dust internal to the
galaxies.  Immediately after the first high-redshift Ly$\alpha$
emitters were finally discovered in the late 1990s~\cite{Huetal1998},
it was realized that they can be used as a probe of reionization: the
neutral IGM can strongly suppress these lines, thus also suppressing
the observed luminosity function \cite{HaimanSpaans1999}.  This field
developed rapidly, both observationally, with the discovery of large
samples of $z\gsim 6$ Ly$\alpha$ emitters (now in the hundreds),
especially in surveys by the Subaru telescope
(e.g. ref.~\cite{Ouchi+2010}).  Theoretical predictions were also
refined, including improved estimates of the impact of absorption on
the observed line profiles, in the presence of a local ionized bubble
around the galaxy, galactic winds causing shifts in the emission line
frequency, and a peculiar velocity of the host
galaxy~\cite{Haiman2002,Santos2004}.  These then begun to be
incorporated into more realistic radiative transfer models through the
inhomogeneous IGM \cite{Dijsktra+2007a}, yielding better estimates of
the (more modest) impact of reionization on the observed luminosity
function \cite{Dijsktra+2007b}.

Finally, as the epoch of reionization receded farther and farther in
redshift, it became increasingly clear that observed galaxies do not
provide sufficient UV radiation to account for this ionization.  The
general search for high--redshift galaxies is therefore an important
part of the history of reionization.  Summarizing this history is
beyond the scope of this article. However, it was not until deep
fields with the Hubble Space Telescope discovered a sizable
population of galaxies that the integrated emission of the observed
objects even came close to providing enough ionizing radiation.  At
the present time, the observed galaxy population at redshift $z\gsim
6$ still fails to reionize the IGM by a factor of a ``few'', unless
extreme assumptions are made about the UV spectrum, and the escape
fraction of ionizing radiation from these galaxies (see,
e.g. \cite{Kuhlen-FG2012}).

\subsection{Reionization in Hierarchical Structure Formation Models}

In parallel with developing observational probes of reionization, over
the past several decades, we have gained an understanding of how
reionization was likely driven by an early generation of stars and
quasars.  As mentioned above, at the current horizon of observations
at $z\sim 7$, the observed population of galaxies fails by only a
factor of $\sim$few to reionize the IGM.  It is quite natural to
attribute the missing ionizing emissivity to fainter galaxies, just
below the current detection threshold.  In support of such an
extrapolation, there is a firm upper limit on the contribution from
faint (individually undetectable) quasars to reionization at $z\sim
6-7$.

A population of black holes at these redshifts ($z\approx 6-7$) would
be accompanied by the copious production of hard ($\gsim$10 keV) X-ray
photons. The resulting hard X-ray background would redshift and would
be observed as a present-day soft X-ray background (SXB).  This
severely limits the abundance of accreting quasar BHs at $z\sim 6-7$:
in order to avoid over-producing the unresolved component of the
observed SXB in the 0.5-2 keV range, these BHs can not significantly
contribute to reionization \cite{DHL2004,SHF2005,McQuinn2012}, or make
up more than a few percent of the present-day total BH mass density
\cite{TanakaHaiman2009,Salvaterra+2012}.  It is important to
emphasize, however, that these constraints still allow accreting BHs
to be dominant over stellar UV radiation at the earliest stages of
reionization $z\sim 15$, partially ``pre-ionizing'' the IGM (see
below).

Because reionization at $z\sim 6-7$ is an (almost) solved problem, the
most interesting open questions concern the earlier stages of
reionization. {\em When did the first light sources turn on?  When did the
IGM first get significantly ionized (and heated)?  What was the
relative contribution of the first stars, of their accreting BH
remnants, or of possibly more exotic sources of ionization, such as
``direct collapse'' supermassive stars or BHs, or decaying dark matter
particles?}

\subsubsection{The Astro-chemistry of ${\rm H_2}$ and The First Stars}

It has long been recognized that the key physics governing the
formation of the first stars (or black holes) is the abundance of
${\rm H_2}$ molecules, which form via gas--phase reactions in the
early universe (in 1967, ref.~\cite{Saslaw+Zipoy1967}).  It is
impossible to form an astrophysical object if gas contracts
adiabatically, because even with the help of cold dark matter, it is
not possible to reach high gas densities. The numerical upper limits
on the gas density in halo cores are extremely tight, especially when
including the entropy generated during adiabatic collapse (see the
recent work in ref.\cite{Visbal+2014}). In the primordial gas, ${\rm
  H_2}$ is the only possible coolant, and determines whether gas can
collapse to high densities.  Following the pioneering paper in 1967 by
Saslaw \& Zipoy~\cite{Saslaw+Zipoy1967}, several groups constructed
complete gas--phase reaction networks, and identified the two possible
ways of forming ${\rm H_2}$ in primordial gas: via the ${\rm H_2^+}$
or ${\rm H^-}$ channels.  These were applied to derive the ${\rm H_2}$
abundance in the smooth background gas in the post--recombination
universe \cite{LeppShull1984}, and also at the higher densities and
temperatures expected in collapsing high--redshift objects
\cite{Hirasawa1969,Matsuda+1969}.

The basic picture that emerged from these early papers is as
follows. The ${\rm H_2}$ fraction after recombination in the
background universe is small ($x_{\rm H2}=n_{\rm H2}/n_{\rm H}\sim
10^{-6}$). At high redshifts ($z\gsim 100$), ${\rm H_2}$ formation is
inhibited even in overdense regions because the required
intermediaries ${\rm H_2^+}$ and H$^-$ are dissociated by the CMB
photons.\footnote{This topic was recently
  revisited~\cite{AlizadehHirata2011} in a more rigorous analysis,
  following the time-dependent, non-equilibrium ${\rm H_2}$ population
  levels. This yielded the same conclusion, i.e. that the
  post-recombination ``intergalactic'' ${\rm H_2}$ abundance is
  negligibly low.}  However, at lower redshifts, when the CMB
temperature drops, a sufficiently large ${\rm H_2}$ abundance builds
up inside collapsed clouds ($x_{\rm H2}\sim 10^{-3}$) at redshifts
$z\lsim 100$ to cause cooling on a timescale shorter than the
dynamical time -- leading to a runaway thermal instability and
eventual star-formation \cite{Hutchins1976,Silk1983,Palla+1983}.  In
summary, these early papers identified the most important reactions
for ${\rm H_2}$ chemistry, and established the key role of ${\rm H_2}$
molecules in cooling the first, relatively metal--free clouds, and
thus in the formation of population III stars.

\subsubsection{The First Stars in Cosmological Structure Formation Models}

The work on ${\rm H_2}$ chemistry was soon connected with cosmological
models for structure formation.  Peebles \&
Dicke~\cite{Peebles+Dicke1968} speculated already in 1968 that
globular clusters, with masses of $\sim 10^{5-6}~{\rm M_\odot}$
(somewhat above the cosmological Jeans mass, set by Compton-heating of
the protogalactic gas by the CMB~\cite{Peebles1965}) forming via ${\rm
  H_2}$ cooling, constitute the first building blocks of subsequent
larger structures.  Early discussions of the formation of galaxies and
clusters have argued that the behavior of gas in a collapsed and
virialized object is determined by its ability to cool radiatively on
a dynamical time~\cite{ReesOstriker1977,WhiteRees1978,DekelSilk1986}.
The same ideas apply on the smaller scales expected for the very first
collapsed clouds~\cite{Silk1977,KashlinskyRees1983}.  Objects that are
unable to cool and radiate away their thermal energy maintain their
pressure support and identity, until they become part of a larger
object via accretion or mergers.  On the other hand, objects that can
radiate efficiently will cool and continue collapsing.

In the late 1990s, these ideas were developed further, in the context
of modern ``bottom-up'' hierarchical structure formation in a
($\Lambda$)CDM cosmology.  In particular, the first DM halos in which
gas can cool efficiently via ${\rm H_2}$ molecules, and condense at
the center, are ``minihalos'' with virial temperatures of $T_{\rm
  vir}\sim$few$\times100$K~\cite{HTL96,Tegmark+1997}.  This is
essentially a gas temperature threshold, above which roto-vibrational
levels of ${\rm H_2}$ are collisionally excited, allowing efficient
cooling.  Because of the emergence of a concordance ($\Lambda$CDM)
cosmology~\cite{WMAP9}, we can securely predict the collapse redshifts
of these minihalos: $2-3\sigma$ peaks of the primordial density field
on the corresponding mass scales of $10^{5-6}~{\rm M_\odot}$ collapse
at redshifts $z=15-20$.\footnote{As an amusing aside: the highest
  redshift in our Hubble volume where we may find a star in a
  collapsed minihalo is $z=65$, corresponding to an $\approx8\sigma$
  fluctuation on the mass scale $10^5~{\rm
    M_\odot}$~\cite{Naoz+2006}.}

\vspace{\baselineskip}
\noindent{\em The Abundance of Low-Mass Minihalos at High Redshift.}

\noindent
The halo mass functions are now robustly determined, since
three--dimensional cosmological simulations reached the required
dynamical range to directly resolve the low--mass end of the high-$z$
halo mass function \cite{Yoshida+2003,Springel+2005}.  The predictions
for the halo mass functions are now therefore limited mainly by the
few $\%$ uncertainty in the normalization ($\sigma_8=0.82\pm 0.02$)
and the power-law index ($n_s=0.972\pm 0.013$) of the primordial power
spectrum \cite{WMAP9}.  A possibly (much) larger source of uncertainty
is that the primordial power spectrum on the relevant scales is not
directly measured - it is extrapolated using the shape of the
processed CDM power spectrum ($P(k)\propto k^{\alpha}$ with
$\alpha\approx -3$ on the relevant small scales).  In principle, the
small-scale power could deviate from this prediction significantly,
reducing the minihalo abundance by a large factor.  This could be
caused by a generic ``running'' ($d\alpha/dk\neq 0$) of the primordial
scalar index \cite{Yoshida+2003b}, or by free-streaming due to the
finite temperature of a low--mass ($\lsim1$ keV) warm dark matter
(WDM) particle \cite{BHO2001,Yoshida+2003}.  While these could have
large effects on the expected halo abundance at $z=15-20$, in
practice, there is no evidence of ``running'' on $\gsim$ Mpc scales,
and the mass of a putative WDM particle is limited to $\gsim1$ keV by
the detections of lensed $z>8$ galaxies \cite{Pacucci+2013} and
gamma-ray bursts \cite{deSouza+2013}.

\vspace{\baselineskip}
\noindent{\em Cosmological Simulations of the Formation of First Stars.}

\noindent
In addition to robustly predicting DM halo formation, high-resolution
3D numerical simulations, including hydrodynamics and ${\rm H_2}$
chemistry, have become possible, with several groups simulating the
cooling and collapse of gas into the first minihalos, located at the
intersections of a ``protogalactic'' cosmic web
\cite{ABN00,BCL02,Yosh+03}.  These simulations showed convergence
toward a gas temperature $T\sim 300~{\rm K}\,$ and density $n \sim
10^{4} \, {\rm cm}^{-3}$, dictated by the thermodynamic properties of
${\rm H_2}$, which allows the collapse of a clump of mass
$10^{2}-10^{3} \,{\rm M_\odot}$ at the center of the high-redshift
minihalos.  These early works suggested that the first stars may have
been unusually massive, a conclusion based on the low mass accretion
rate in the cores of these halos.  In a self-gravitating gas, the mass
accretion rate depends only on the sound speed $c_s$, and is of order
$\sim c_s^3/G\propto T^{3/2}/G$ (e.g.~\cite{Shu77}).
Three-dimensional simulations have confirmed this scaling
(e.g.~\cite{ON07,WTA08,SBH10}), and in minihalos, the corresponding
mass accretion rates are $\sim 10^{-3} {\rm M_\odot~yr^{-1}}$.  At
this accretion rate, the mass that will accumulate in the halo nucleus
within a Kelvin-Helmholtz time ($\sim 10^5$ years; only weakly
dependent on mass for massive protostars) is of order $10^2~{\rm
  M_\odot}$.

Simulations in the past few years have been pushed to higher spatial
resolution, and, in some cases with the help of sink particles, were
able to continue their runs beyond the point at which the first
ultra-dense clump developed.  The gas in the central regions of at
least some of the early minihalos were found to fragment into two or
more distinct clumps
\cite{Turk+09,Stacy+10,Greif+11,Clark+11,Prieto+11}. This raises the
possibility that the first stars formed in multiple systems, and that
some of these stars had masses $\lsim 100~{\rm M_\odot}$, lower than
previously thought (but see \cite{Turk+12} for still higher resolution
simulations that suggest less efficient fragmentation).

\vspace{\baselineskip}
\noindent{\em The First Stars and the Beginning of Reionization.}

\noindent
Even if star--formation in minihalos was inefficient, these early
minihalos should have begun ionizing the universe.  With a usual
Salpeter IMF, each proton in a population of stars would create
$\approx 4,000$ ionizing photons (e.g. \cite{HL97}).  A population of
massive, metal--free stars would increase the efficiency of ionizing
photon production per unit mass by a factor of $\sim 20$, to $\sim
10^5$~\cite{TS2000,BKL2001,Schaerer2002}. Each proton accreted onto a
BH could release $\sim0.1 m_p c^2=0.1$GeV of energy, most of it in
ionizing radiation, implying enough energy to cause up to $10^7$
ionizations.  These numbers suggest that once a small fraction ($\lsim
10^{-5}$) of the gas in the universe is converted into massive stars
or black holes, a significant ionization of the rest of the IGM can occur.

The simple argument above ignores recombinations (in a fully ionized
IGM, each hydrogen atom would recombine several times at $z\gsim 10$)
and the details of the ionizing spectrum and the photoionization
process (which, in the case of hard-spectrum sources, needs to account
for secondary ionizations by photoelectrons). Nevertheless, the main
conclusion, namely that early stars or black holes should have
``kick-started'' reionization, is hard to avoid. In particular, if
each minihalo is allowed to form PopIII stars, it would result in a
significant $\tau_e$, in tension with the electron scattering optical
depth measured by \texttt{WMAP} and
\texttt{Planck}~\cite{HaimanBryan2006,Visbal+2015b}.  Indeed, in the wake of the
``false alarm'' from WMAP's first measurement of a large $\tau_e$,
several authors investigated the even more efficient
``pre-ionization'' of the IGM at $z\sim 20$ by accreting BHs
\cite{Madau+2004, RicottiOstriker2004}.  While those models with a
large X-ray emissivity are now ruled out, a contribution from
early accreting BHs still remains a natural possibility, especially if
fragmentation in early halos (mentioned above) leads to the frequent
formation of high-mass X-ray binaries \cite{MFS2013,Jeon+2014,TPH2012}.

\subsubsection{Global Reionization Models in a Hierarchical Cosmology}

Beginning in the late 1990s, detailed models were put together, in
which the well-understood cosmological dark matter halos were
populated by stars or black holes (early examples
include~\cite{Shapiro+1994,HL97,HL98}).  These models allowed
physically motivated calculations of the entire reionization history,
between $6\lsim z\lsim 30$, to be confronted with data.

An important physical ingredient in reionization models, especially at
the earliest stages, is global radiative feedback.  Soon after the
first stars appear, early radiation backgrounds begin to build up,
resulting in feedback on subsequent star--formation.  In particular,
the UV radiation in the Lyman--Werner (LW) bands of ${\rm H_2}$ can
photodissociate these molecules and suppress gas cooling, slowing down the global star-formation rate
\cite{HRL97,ON99,HAR00,CFA00,MBA01,Ricotti+01,
  Ricotti+02b,MBH06,WA07,ON08,JGB08,WA08a,WA08b,Whalen+08,MBH09}.

If the metal--free stars forming in the early minihalos were indeed
very massive ($\sim 100~{\rm M_\odot}$), then these stars would leave
behind remnant BHs with similar masses \cite{Heger+03}, and could
produce significant X-rays, either by direct accretion or by forming
high-mass X-ray binaries.  A soft X-ray background at photon energies
of $\gsim 1$keV, at which the early intergalactic medium (IGM) is
optically thin, then provides further global feedback: both by heating
the IGM, and by catalyzing ${\rm H_2}$ formation in collapsing halos
\cite{HRL96,Oh01,Venkatesan+2001,GloverBrand2003,Madau+2004,ChenJordi2004,Ricotti+2005,Mirabel+2011}.

On the other hand, if fragmentation was very efficient, and the
typical PopIII stars had low masses, they would not leave BH remnants
and they would have softer spectra, with copious infrared (IR)
radiation at photon energies $\sim 1$eV.  Similar to LW and X--ray
photons, these photons have a mean--free path comparable to the Hubble
distance, building up an early IR background.  If soft--spectrum
stars, with masses of a few ${\rm M_\odot}$, contributed $\gsim 0.3\%$
of the UV background (or their mass fraction exceeded $\sim 80\%$),
then their IR radiation would have dominated the global (negative)
radiative feedback in the early Universe \cite{JemmaHaiman2012}. This
feedback is different from the LW feedback from high-mass stars, and
occurs through the photo-detachment of ${\rm H^-}$ ions, necessary for
efficient ${\rm H_2}$ formation.  Nevertheless, the baryon fraction
which must be incorporated into low--mass stars in order to suppress
${\rm H_2}$--cooling is comparable to the case of high-mass stars.

The net effect of the above long-range ``global'' feedback effects
remains poorly understood. This is a significant outstanding question,
as these feedback effects likely determined the earliest stages of the
global reionization history.  The difficulties with a self-consistent
reionization model are two-fold. First, one needs a detailed ab-initio
understanding of the feedback on individual protogalaxies with
different masses and redshifts.  Second, the feedback processes (such
as photo--ionization heating, ${\rm
  H_2}$--dissociation~\cite{Dijkstra+2008,Ahn+2009}, and also
metal--enrichment), are all affected by the strong clustering of the
earliest sources.  Semi-analytical models have included {\it either}
various feedback effects
(e.g. \cite{HL98,HaimanHolder2003,WL2003,Cen2003,Iliev+2007,Johnson+2007})
{\it or} the effect of source clustering on the HII bubble--size
distribution (e.g. \cite{FZH2004}), but not yet both self-consistently.
Only the first steps were taken towards such a self-consistent
treatment, incorporating photo-ionization feedback, in a simplified
way, into a model that partially captures the source clustering (only
in the radial direction away from sources)~\cite{Kramer+2006}.

Numerical simulations do not have the dynamical range for an ab-initio
treatment of this issue.  The minihalos hosting the first stars arise
from primordial perturbations on the scale of $\sim 10~{\rm (comoving)
  kpc}$.  On the other hand, the global feedback effects operate over
a distance comparable to the Hubble length, $\sim 1~{\rm Gpc}$.  Even
if one were to resolve a minihalo with only $10^3$ particles, 3D
simulations would need to cover a factor of $\sim 10^6$ in spatial
scales (or contain $10^{18}$ particles).  Clearly, this can not be
achieved by N-body simulations - let alone hydrodynamical simulations
that include the basic physics, such as cooling, chemistry, and
radiative transfer.\footnote{For reference, the largest existing
  N-body simulation is the Millennium-XXL project with
  $3\times10^{11}$ particles.}  Semi-numerical treatments
\cite{MF2007} can offer an order of magnitude higher dynamical range,
and have incorporated radiative feedback \cite{Zahn+2011}, but are
still short of covering the required range of scales (i.e. still to
need to prescribe small-scale non-linear processes with sub-grid
prescriptions).

\subsubsection{Stars vs. Black Holes as Sources of Reionization}

As is clear from above, whether the first stars were formed as single
stars, or in binaries, matters for the early stages of reionization.
If the majority of the first stars formed high--mass X-ray binaries,
they could have produced sufficient X--rays to significantly change
the expected ``Swiss--cheese'' morphology of
reionization\cite{Oh01,Venkatesan+2001,GloverBrand2003,ChenJordi2004}.
The thickness of the edges of the cosmological ionized regions would
be of order the mean free path of the typical ionizing photon.  For
the UV photons from stars, this mean free path is small, resulting in
sharply defined ionization fronts.  But for the hard spectra of X-ray
binaries (or more generally, accreting black holes), the mean free
path can be long, comparable to the Hubble distance for photon
energies above $E>[(1+z)/11]^{1/2} x_{\rm HI}^{1/3}$keV (where $x_{\rm
  HI}$ is the mean neutral H fraction in the IGM).  The diffuse nature
of the boundaries of individual ionized regions could be detectable,
in principle, through 21cm or Ly$\alpha$
observations~\cite{KramerHaiman2008,ThomasZaroubi2008}.

Since X-rays in the early Universe can travel across the Hubble
distances, they can also change the global reionization topology. The
X-rays would ionize and heat the plasma much more uniformly than stars
(although they could increase the ionized fraction only to $\sim20\%$;
nearly all of the energy of the fast photo-electron from the first
ionization will subsequently go into heating the IGM).  If X--rays are
sufficiently prevalent, a range of other interesting effects will
occur: the extra heating will raise the pressure of the plasma
everywhere, making it resistant to clumping, and more difficult to
compress to form new galaxies~\cite{Oh01,TanakaHaiman2009}.  On the
other hand, X--rays can penetrate the successfully collapsing
protogalaxies and can ionize hydrogen and helium atoms in their
interior. This will catalyze the formation of molecular hydrogen, and
help the gas to cool and form new stars~\cite{HAR00}. These effects
will leave behind their signatures in the spatial distribution of
neutral and ionized hydrogen and helium in the
Universe. Distinguishing these different global morphologies could be
possible in 21cm experiments~\cite{FOB2006}, or in the CMB through the
kSZ effect \cite{Mesinger+2012}.

There are other possible sources of X--rays, in addition to binaries,
connected to the formation of the first stars. One example is gas
accretion onto the black--hole remnants left behind by the collapse of
single (super)massive
stars\cite{Heger+03,Madau+2004,RicottiOstriker2004}. Another possible
source is supernovae (SNe): if the first stars exploded as SNe, then
similar X--rays would be produced by thermal emission from the gas
heated by these SNe, and by the collisions between the energetic
electrons produced in the SN explosion and the CMB
photons~\cite{Oh01}.  Thermal emission from a hot ISM has indeed been
found to dominate the soft X-ray emission in a sample of local
star-forming galaxies~\cite{Mineo+2012}.

We emphasize that X-ray sources can not contribute significantly to
reionization at lower redshifts, as they would then have overproduced
the unresolved X-ray background \cite{DHL2004,SHF2005,McQuinn2012},
nor could they have elevated the ionized fraction to $\gsim 20\%$ at
early times.  However, a smooth partial ``pre''ionization by sources
whose spectrum peaks near $\sim 1$keV remains a plausible an
interesting scenario.

In summary -- the simplest possibility is that the first stars and
black holes started reionizing the universe by redshift $z\approx
15-25$; the process then was completed predominantly by small
galaxies, in the redshift range $6\lsim z \lsim
10$.\footnote{Reionization must end by $z\sim 6$, as shown recently
  using the fraction of dark Ly$\alpha$ and Ly$\beta$ pixels in a
  sample of 22 quasars~\cite{McGreer+2015}}. The relative contribution
of these two types of sources is yet to be understood, especially at
the earliest epochs; as is the net effect of the global radiation
backgrounds that should build up early on.  These are fundamental
outstanding questions. The relative abundance of the two types of
sources determined the global ionization topology, and their feedback
processes likely drove global time-evolution of reionization.

Finally, for completeness, it is useful to note that there are several
other, more exotic sources that may have contributed to reionization
in principle. These include decay products of various different dark
matter
particles~\cite{ChenKamion2004,HansenHaiman2004,Kasuya+2004,BK2006,RMF2007},
high energy cosmic rays~\cite{Oh01,SV2004,SB2007}, or excess
small-scale structure formation arising from primordial
non-Gaussianities~\cite{Chen+2003}, a running of the spectral
index~\cite{Yoshida+2003b}, or a red spectral
tilt~\cite{Alvarez+2006,MortonsonHu2008}.  Many of these alternatives
were proposed in the wake of the anomalously high $\tau_e$ in the
WMAP3 data, and, at the present time, there is no longer a need for
these additional contributions.

\section{Can We Detect the First Stars Directly?}
\label{haimansec:4}

As mentioned in the Introduction, reionization is a probe of the
earliest light sources.  The redshift and duration of reionization of
reionization, inferred from quasar absorption spectra, 21cm
signatures, and the CMB, will place a constraint on the host halos and
the ionizing efficiencies.  The observed level of ``patchiness'' will
constrain the spectral hardness of the typical source, constrain
the relative contribution of stars and black holes, and shed light on
the birth and death of the first galaxies.

One may, however, ask: is this the best we can do, or is there a hope
to directly detect the light from the first stars or black holes?  It
is simple to obtain a rough estimate for the stellar mass of in a
proto-galaxy, or the mass of a bright (near-Eddington) black hole,
which could be detected at the $\sim 1$nJy detection threshold in a
deep exposure with the {\em James Webb Space Telescope}.  At $z=10$,
this requires a mass of about $10^5~{\rm M_\odot}$, either in stars
\cite{HL97} or in a BH~\cite{HL98}.  (The former is consistent with a
recent detailed estimate~\cite{Zackrisson+2011}.)  It is quite
plausible (or even likely) that the {\em very first} galaxies and
quasars were below this threshold.

So what hopes do we have of directly seeing the light of these first
sources?  I believe there are three possibilities.

First, observations can be about an order of magnitude more sensitive,
using a foreground cluster to gravitationally lens and magnify the
$z\sim 10$ background sources.  Indeed, there are two examples of
detecting $z=8-10$ galaxies \cite{Zheng+2012,Coe+2013} using this
technique on 28 foreground clusters the CLASH survey
\cite{Postman+2012}.  The on-going Hubble Frontier Fields, going an
order of magnitude deeper using 4-6 clusters.  This technique gives a
chance of discovering $10^4~{\rm M_\odot}$ mini-galaxies or miniquasars
at $z\sim 10$.

Second, and most promising, would be to detect the individual
supernovae (SNe) from the first stellar populations.  Even ``normal''
core collapse SNe are bright enough to be visible well beyond $z=10$,
and the pair instability SNe expected from massive PopIII stars with
$\sim 130-250~{\rm M_\odot}$ would be even brighter.  It has been
shown that {\em JWST} could detect many hundreds of these SNe; the
challenge will be that repeated observations will be required on many
{\em JWST} fields, separated by years, to identify the slowly evolving
light-curves of these ultra-distant SNe \cite{MJH2006}.

Third, even if we cannot directly detect individual stars, black
holes, or SNe, we can still directly detect their cumulative faint
emission, through the technique known as ``intensity mapping''.  In
general this technique consists of ``tomographic'' observations of the
fluctuating intensity in the emission lines from faint, individually
undetectable sources~\cite{Righi+2008,Visbal+Loeb2010}.  In practice,
at least two emission lines are required, so that their spatial
fluctuations (in sky position and in redshift space) can be
cross--correlated, eliminating contaminating signals from a foreground
line.  The same technique can be applied, in principle, to the strong
HeII~1640\AA\ emission lines expected from the first generation PopIII
stars, cross--correlated with CO emission from the same galaxies, or
with 21cm emission from the IGM \cite{Visbal+2015}.  This would
require a next-generation UV instrument (the example considered in
\cite{Visbal+2015} is a space-borne 2m dish, with 100 individual
detector pixels with spectral resolution R=1000).

\section{The Future}

As the rest of this book will make clear, the future is bright, with
\texttt{JWST}, \texttt{ALMA}, and several new 21cm experiments coming
on line, allowing us to peer farther back in redshift.  The main
challenge will likely become to constrain parametric models, since it
is unlikely that we will have full, ab-initio calculations of the
reionization process incorporating all the relevant physics, on
scales ranging from star-formation inside minihalos, to the global
radiative feedback processes operating on the Hubble scale.  With a
combination of multiple observational probes, this will nevertheless
give us a chance to understand the cosmic history of structure
formation from its very beginning.

\begin{acknowledgement}

I thank my students and collaborators, who taught me a lot about
reionization, the US federal agencies NASA and NSF for funding much of
my research, and Andrei Mesinger for the initiative to put together
this volume, and his patience and dedication during the production
process.

\end{acknowledgement}

\bibliographystyle{unsrt}

\end{document}